\documentclass[sn-aps]{sn-jnl}

\jyear{2021}%

\theoremstyle{thmstyleone}%
\newtheorem{theorem}{Theorem}
\theoremstyle{thmstyletwo}%

\theoremstyle{thmstylethree}%

\begin{document}

\title[Low-Overhead Code Concatenation Approaches for Universal QC]{Low-Overhead Code Concatenation Approaches for Universal Quantum Computation}


\author*[1,2]{\fnm{Eesa} \sur{Nikahd}}\email{nikahd@aut.ac.ir}

\author[2]{\fnm{Morteza} \sur{Saheb Zamani}}\email{szamani@aut.ac.ir}

\author[2]{\fnm{Mehdi} \sur{Sedighi}}\email{msedighi@aut.ac.ir}
\affil[1]{\orgdiv{Computer Engineering Department}, \orgname{Shahid Rajaee Teacher Training University}, \orgaddress{\city{Tehran}, \country{Iran}}}
\affil[2]{\orgdiv{Quantum Design Automation Lab}, \orgname{Amirkabir University of Technology}, \orgaddress{\city{Tehran}, \country{Iran}}}


\abstract{Fault-tolerant quantum circuit design can be done by using a set of transversal gates. However, as there is no quantum error correction code with a universal set of transversal gates, several approaches have been proposed which, in combination of transversal gates, make universal fault-tolerant quantum computation possible. Magic state distillation, gauge fixing, code switching, code concatenation and pieceable fault-tolerance are well-known examples of such approaches. However, the overhead of these approaches is one of the main bottlenecks for large-scale quantum computation. In this paper, two approaches for universal fault-tolerant quantum computation, mainly based on code concatenation, are proposed. The first approach combines the code concatenation approach with code switching, pieceable fault-tolerance or magic state distillation and the second approach extends the nonuniformity of the concatenated codes by allowing to apply CNOT gates between different codes. The proposed approaches outperform the code concatenation approach in terms of both number of qubits and code distance and have also significantly less resource overhead than magic state distillation. This is achieved at the cost of reducing the effective distance of the concatenated code for implementing non-transversal gates.}

\keywords{Fault-tolerant Quantum Computation, Universal Quantum Computation, Quantum Error Correction}



\maketitle

\section{Introduction}\label{sec:intro}
Quantum computers have the potential to efficiently solve certain problems such as integer factorization \cite{ahsan2015architecture7}\cite{shor1994algorithms1} and simulation of quantum systems \cite{zalka1998efficient2} which are prohibitively time-consuming using classical computers. This computational advantage of quantum computers comes from the unique quantum mechanical properties such as superposition and entanglement, which have no classical analogue \cite{nielsen2010quantum3}.

Quantum bits or qubits are the fundamental units of information in quantum computing. Unfortunately, qubits are fragile and tend to lose their information due to the environmental noise resulting in decoherence \cite{nielsen2010quantum3}\cite{unruh1995maintaining4}. Furthermore, the physical implementations of quantum operations in any technology are imperfect \cite{unruh1995maintaining4}\cite{mazzola2010sudden5}. Quantum noise, due to decoherency of quantum states and imperfect quantum operations, is the most important challenge in constructing large-scale quantum computers \cite{nielsen2010quantum3}\cite{metodi2006quantum6}\cite{ahsan2015architecture7}.

The most common approach to cope with this challenge is the use of quantum error correction codes and fault-tolerant operations to perform quantum computation. In this approach, a logical qubit is encoded into multiple physical qubits using a suitable error correction code. Logical operations are applied directly on the encoded qubits in such a manner that decoding is not required. After that, if a qubit becomes erroneous, that qubit can be corrected using application of the quantum error correction procedure. The logical operations can potentially spread errors due to the interactions among qubits. Therefore, to preserve the veracity of computation, these operations must be implemented fault-tolerantly in such a way that they do not propagate errors from a corrupted qubit to multiple qubits in a codeword.


$Transversal$ implementation of logical gates is widely considered as a simple and efficient method for fault-tolerant quantum computation (FTQC) \cite{shor1996fault9}\cite{anderson2014fault10}, where a transversal gate refers to a gate which does not couple qubits inside a codeword. Unfortunately, there is no quantum code with a universal set of transversal gates \cite{eastin2009restrictions11}. Several approaches have been proposed which, in combination with transversal gates, make universal FTQC possible. Magic state distillation \cite{bravyi2005universal12}, gauge fixing \cite{paetznick2013universal17}\cite{bombin2015gauge18}, pieceable fault-tolerance \cite{yoder2016universal21}, code switching \cite{anderson2014fault10}\cite{stephens2008asymmetric13}\cite{choi2015dual14} and code concatenation \cite{jochym2014using15}\cite{nikahd2016non16} are well-known examples of such approaches.

Magic state distillation (MSD) is a procedure which uses only Clifford operations to increase the fidelity of non-stabilizer states that can be used to realize non-Clifford gates. This procedure is orders of magnitude more costly than transversal gates and can incur a significant resource overhead for the implementation of a quantum computer \cite{jochym2014using15}\cite{fowler2012surface19}\cite{chamberland2019fault}\cite{chamberland2020very}\cite{haah2018codes}\cite{hastings2018distillation}.

Gauge fixing is an alternative approach for universal quantum computation without the need for a special ancillary state prepared by MSD, using only one quantum code. The method proposed by Paetznick and Reichardt \cite{paetznick2013universal17} is an example of this approach. This method has been described based on the [[15, 7, 3]] quantum Hamming code as an example. In this code, by considering the first logical qubit as data qubit and fixing the other six logical qubits into the encoded $\vert 0^{\otimes6}\rangle$
state as gauge qubits, CCZ gate will be transversal. Applying the transversal $H$ gate to all of the 15 qubits of the code performs a logical $H$ to the first logical qubit. However, as its application corrupts the state of the gauge qubits, additional error correction and transversal measurements are needed in order to restore their state into $\vert 0^{\otimes6}\rangle$.

Yoder et al.\cite{yoder2016universal21}\cite{yoder2018practical} proposed a novel approach to overcome the limits of non-transversality, namely pieceable fault-tolerance (PFT). In this approach, a non-transversal circuit is broken into fault-tolerant pieces with rounds of intermediate error corrections in between to correct errors before they propagate to a set of non-correctable errors. As an example, fault-tolerant implementation of CCZ was developed for the 7-qubit Steane code. This considerably reduces the resource overhead in comparison with MSD. However, this approach is unable to find a fault-tolerant structure for non-transversal single-qubit gates, such as $T$, without additional ancillae. 

The code switching method achieves a universal set of transversal gates by switching between two different quantum codes $C_1$ and $C_2$ where the non-transversal gates on $C_1 (C_2)$ have transversal implementations on $C_2 (C_1)$. To this end, a fault-tolerant switching network is required to convert $C_1$ into $C_2$ and vice versa. A general approach to convert codes uses teleportation \cite{choi2015dual14}\cite{oskin2002practical20}. Alternatively, some methods have been proposed for direct fault-tolerant conversion between the codes of Reed-Muller code family \cite{anderson2014fault10}\cite{quan2018fault}. Moreover, a method has been published in \cite{yoder2016universal21} using pieceably fault-tolerant gates. In \cite{beverland2021cost}, code switching is used to implement $T$ gate in color codes. 

Uniform code concatenation method \cite{jochym2014using15} uses two different quantum codes, namely $C_1$ and $C_2$, in concatenation to achieve universal fault tolerance. In this approach, a logical qubit is encoded into $C_1$ and then each qubit of $C_1$ is in turn encoded into $C_2$, where for each non-transversal gate $U$ on $C_1$, there is an equivalent transversal implementation on $C_2$. In our previous work \cite{nikahd2016non16}, a method called nonuniform code concatenation has been proposed which outperforms the uniform one. The main idea behind this approach is that the non-transversal implementation of gate $U$ does not necessarily involve all of the $C_1$ qubits and it is shown that it is only needed to encode the involved qubits in the second level of concatenation. Although these approaches eliminate the need for magic state distillation, the number of necessary physical qubits to code the logical information is large. Moreover, for the non-transversal gates on $C_1$ or $C_2$, the overall distance of the concatenated code is sacrificed.

\begin{figure}
	\centering
	\includegraphics[trim=0in 0in 0in 0in,clip,width=0.7\columnwidth]{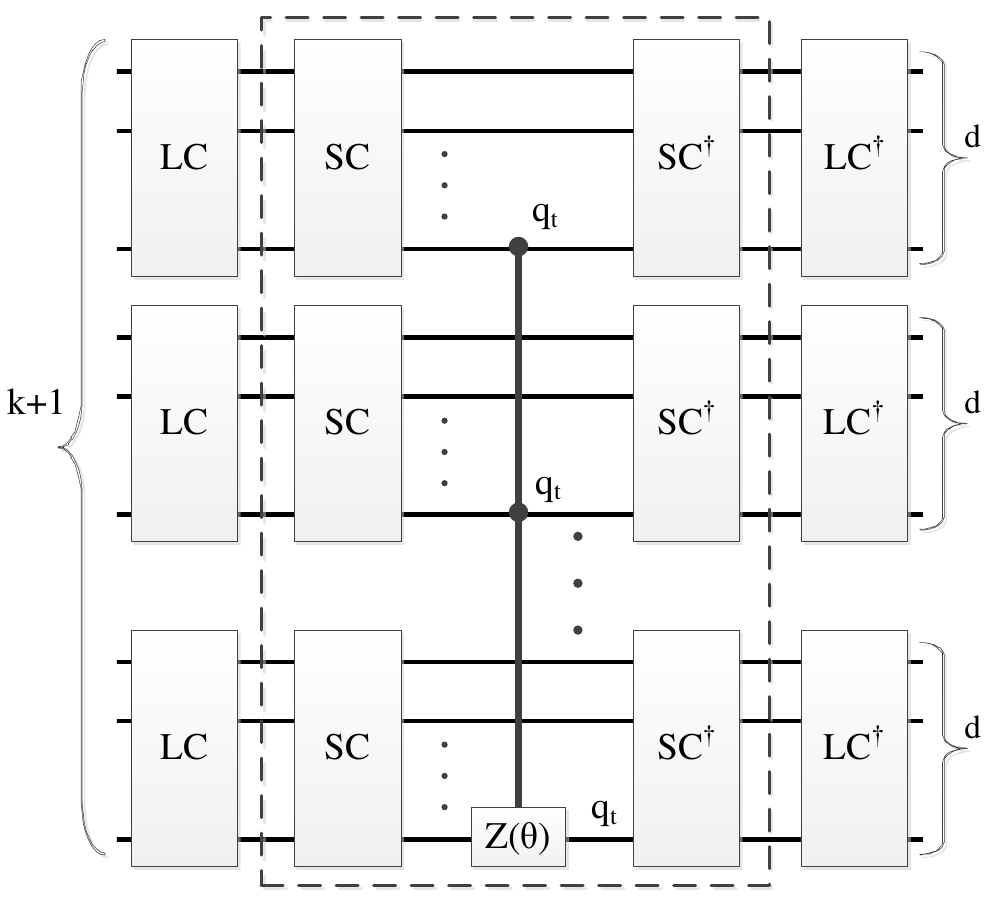}
	\caption{Non-transversal application of logical $C^{k}Z(\theta)$ gate for a stabilizer code [[n, 1, d]] by involving only $d$ qubits of each code block and only one physical $C^{k}Z(\theta)$, where $d$ is the code distance \cite{nikahd2016non16}. $SC$ is an abbreviation for staircase of $CNOT$s and $LC$ is a circuit containing only local Clifford gates. Note that only the active qubits are shown.}
	\label{fig:CkZ}
\end{figure}

The main motivation of the proposed approaches in this paper is as follows. In \cite{nikahd2016non16} it was shown that:
\begin{theorem}
	For a stabilizer code $[[n, 1, d]]$, a logical $C^{k}Z(\theta)$ gate can be implemented non-transversally using a circuit that involves only $d$ qubits, called \textit{active qubits}, of each code block as shown in Fig. \ref{fig:CkZ}, where $Z(\theta)=diag(1,exp^{i\theta})$ and $k\in\{0, 1, 2, ...\}$.
	\label{theorem:CkZ}
\end{theorem}

Magnifying this circuit draws our attention to the only possible non-Clifford gate $C^kZ(\theta)$, which is often the main challenge for fault-tolerant implementation of the circuit. In the uniform and non-unifrom code concatenation approaches, both $C^kZ(\theta)$ and Clifford gates in the circuit must be transversal on $C_2$. Here, two related important questions arise: 
1. Instead of using two codes $C_1$ and $C_2$ with a complementary set of transversal gates, is it possible to use a more efficient code $C_2'$ than $C_2$ in the second level of concatenation with possibly non-transversal implementation for $C^kZ(\theta)$ and exploit other methods to implement it fault-tolerantly on $C_2'$?
2. Is it possible to encode only the target qubit of $C^kZ(\theta)$ of each codeword, i.e., $q_t$, using a code with transversal $C^kZ(\theta)$ and encode the rest of active qubits using a more efficient code with possibly non-transversal implementation for $C^kZ(\theta)$?
Fortunately the answer to both questions is yes and this leads to two approaches that we call \textit{Hybrid Code Concatenation (HCC)} and \textit{Extended NonUniform Code Concatenation (ENUCC)}, respectively. Describing how these approaches are accomplished is the main focus of this paper.

\section{HCC: Hybrid code concatenation} \label{sec:proposed1}
In this section, a hybrid approach is proposed which combines the code concatenation approach \cite{jochym2014using15}\cite{nikahd2016non16} with code switching \cite{anderson2014fault10}\cite{stephens2008asymmetric13}, PFT \cite{yoder2016universal21} or MSD \cite{bravyi2005universal12}, to provide a low-overhead universal fault-tolerant scheme. The proposed method encodes the information using $C_1$ in the first level of concatenation and then the qubits of $C_1$ are in turn encoded into $C_2$ code, either uniformly or nonuniformly. As there is no quantum code with a universal set of transversal gates, there is at least one non-transversal gate $U$ on $C_1$. Let the non-transversal implementation of $U$ on $C_1$ be constructed using a set of gates $G=\{g_1, g_2, ..., g_m\}$. In the proposed approach, there may exist some gates $g_{nt} \in G$ with non-transversal implementation on $C_2$. This is in contrast to both uniform and nonunifrom code concatenation approaches where all of the $G$ gates must be transversal on $C_2$. Indeed, the proposed method uses a more efficient code than code concatenation approaches in the second level of concatenation but with the overhead of using more costly approaches such as code switching, MSD or PFT for applying the non-transversal gates $g_{nt}$ on $C_2$, fault-tolerantly. The idea behind this method is that the number of non-transversal gates $g_{nt}$ on $C_2$ may be relatively small.

Based on the implementation of the non-transversal gate $U$, the qubits of $C_1$ can be partitioned into two separate sets, namely \textit{active} and \textit{non-active} qubits. Active qubits are the coupled qubits and the remaining qubits belong to the set of non-active qubits. In the proposed approach, the active qubits should be encoded using $C_2$ in the second level of concatenation whereas the non-active qubits can be left unencoded, encoded using $C_1$ or encoded using $C_2$. We refer to these three cases in dealing with the non-active qubits as Case I, Case II and Case III, respectively. The ways in which the gates are applied in the proposed approach are as follows.

The shared transversal gates on both $C_1$ and $C_2$ are globally transversal on the concatenated code and are therefore, fault-tolerant. For the transversal gates on $C_1$ with non-transversal implementation on $C_2$, although a single physical error may corrupt a $C_2$ logical qubit, it can be corrected using error correction procedure on $C_1$, similar to the code concatenation approaches. 

The main challenge is fault-tolerant application of the non-transversal gates on $C_1$ referred to as $U$. As mentioned above, the implementation of $U$ on $C_1$ uses some gates $G=\{g_1, g_2, ..., g_m\}$. These gates can be partitioned into two non-overlapping sets, namely $S_t$ and $S_{nt}$. A gate $g_i$ belongs to $S_t$ if it has a transversal implementation on $C_2$ and belongs to $S_{nt}$, otherwise. The gates of $S_t$ are transversal and therefore, fault-tolerant in the second level of concatenation. However, the proposed method exploits other existing approaches such as code switching, MSD or PFT for fault-tolerant application of the $S_{nt}$ gates as they are non-transversal on $C_2$. Therefore, each gate $g_i\in G$ is applied fault-tolerantly in the second level and a single error on one of the active qubits causes only a single physical error in each of them which are themselves encoded blocks of $C_2$. Consequently, this error can be corrected using error correction procedure on $C_2$. 

This approach can lead to a low-overhead fault-tolerant implementation of the non-transversal gate $U$ if the number of non-transversal gates $g_{nt}$ (i.e., $\vert S_{nt} \vert$) is relatively small for the selected codes $C_1$ and $C_2$. Fortunately, as stated in Theorem \ref{theorem:CkZ}, for a stabilizer code $[[n, 1, d]]$, a logical $C^{k}Z(\theta)$ gate can be implemented using some local Clifford ($LC$) and CNOT gates on each codeword and only one physical $C^{k}Z(\theta)$ gate. Therefore, for a non-transversal $C^{k}Z(\theta)$ on both $C_1$ and $C_2$, we have $\vert S_{nt}\vert =1$ where $C_2$ is a CSS code, i.e., it has transversal implementation for CNOT gate. It is worth noting that the $LC$ gates will not be a challenge even if they are non-transversal on $C_2$. This is because one can simply replace the $C_1$ code with $C'_1$, where $C'_1$ is a code generated by applying $LC$ gates on $C_1$. While $C'_1$ has the same properties as $C_1$, the logical $C^{k}Z(\theta)$ can be applied on $C'_1$ as shown in the dashed box of Fig. \ref{fig:CkZ} without the need for applying $LC$ gates.

Let us now describe the HCC method in detail through some examples using the 5-qubit perfect code and 7-qubit Steane code. Although the following examples are described based on the combination of code concatenation and MSD in two levels of concatenation, one can easily replace MSD with code switching or PFT with no considerable modification and also generalize it for higher levels of concatenation.

\subsection{HCC-based code examples using the Steane code as $C_1$} \label{subsec:steaneCode_HCC}
\begin{figure}
	\centering
	\includegraphics[trim=0in 0in 0in 0in,clip,width=0.6\columnwidth, height=0.3\textheight]{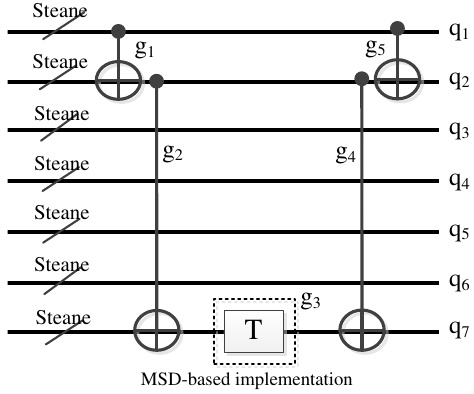}
	\caption{Fault-tolerant implementation of $T$ based on the HCC approach for the 49-qubit code.}
	\label{fig:steaneT_HCC}
\end{figure}

The Steane code is considered as $C_1$ in this section. The Clifford set \{$H$, $S=C^{0}Z(\frac{\pi}{2})$, $CZ=C^{1}Z(\pi)$\} along with a non-Clifford gate such as $T=C^{0}Z(\frac{\pi}{4})$ construct a universal set. Clifford gates are transversal on Steane while $T$ is not. The non-transversal implementation of $T$ on a Steane code block consists of one $T$ and four CNOT gates as shown in Fig. \ref{fig:steaneT_HCC}. Thus, active qubits are $\{q_1, q_2,q_7\}$ and non-active qubits are $\{q_3, q_4, q_5, q_6\}$. By choosing the Steane code as $C_2$, we have $S_{nt}=\{g_3\}$ and $S_t=\{g_1, g_2, g_4, g_5\}$. The active qubits are encoded using the Steane code and based on whether the non-active qubits are left unencoded or are encoded using Steane, a 25- or 49-qubit code is produced, respectively.

Clifford gates are transversal in both levels of coding hierarchy and are thus, fault-tolerant for the proposed concatenated codes. For fault-tolerant implementation of the logical $T$ gate, the gates of $S_t$ are applied transversally on Steane ($C_2$) and the $T$ gate ($g_3$) can be applied using MSD approach, as shown in Fig. \ref{fig:steaneT_HCC}.

\subsection{HCC-based code examples using the 5-qubit code as $C_1$} \label{subsec:5qubitCode_HCC}
\begin{figure}
	\centering
	\includegraphics[trim=0in 0in 0in 0in,clip,width=0.7\columnwidth, height=0.4\textheight]{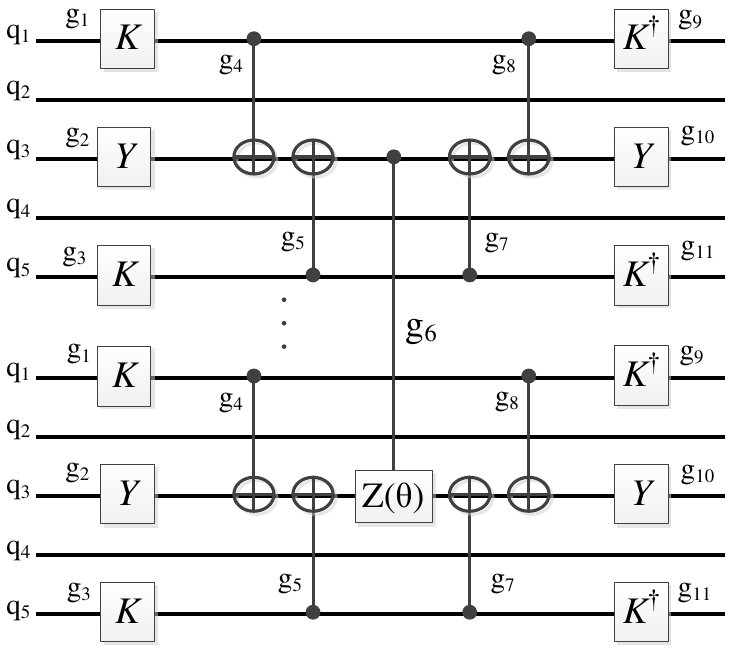}
	\caption{Non-transversal implementation of $C^{k}Z(\theta)$ for the 5-qubit code.}
	\label{fig:CKZ5}
\end{figure}

In this section, the 5-qubit code is selected as $C_1$ and a logical qubit is encoded into the 5-qubit code in the first level of concatenation. Let $M$ be $\{T=C^{0}Z(\frac{\pi}{4})$, $S=C^{0}Z(\frac{\pi}{2})$, $CZ=C^{1}Z(\pi)\}$. The gates of $M$ along with $K=SH$ form a universal set for quantum computation. The $K$ gate is transversal on the 5-qubit code but the gates of $M$ are not. The gates of $M$ belong to the class of $C^{k}Z(\theta)$ gates and thus, as described before, a CSS code such as Steane can be selected as $C_2$ with $\vert S_{nt}\vert =1$. Based on Fig. \ref{fig:CKZ5}, which shows the non-transversal implementation of $M$ gates on the 5-qubit code, we have active qubits as $\{q_1, q_3, q_5\}$ and non-active qubits as $\{q_2, q_4\}$ and also $S_{nt}=\{g_6\}$ and $S_t=\{g_1, g_2, g_3, g_4, g_5, g_7, g_8, g_9, g_{10}, g_{11}\}$ only for the $T$ gate (note that $S$ and $CZ$ are transversal on Steane). The active qubits are encoded using Steane and the non-active qubits can be either left unencoded, encoded using the 5-qubit code or encoded using Steane which leads to a 23-, 31-, or 35-qubit code, respectively.

The $K$ gate can be applied transversally for all of the 23-, 31- and 35-qubit codes, as it is transversal on both the 5-qubit and Steane codes. The $S$ and $CZ$ gates are transversal on Steane and therefore, they can be applied fault-tolerantly on the concatenated code without need to use MSD. However, for fault-tolerant implementation of $T$, the proposed method uses MSD for applying $g_6$ ($T$). 

\section{ENUCC: Extended nonuniform code concatenation} \label{sec:proposed2}
In this approach, a logical qubit is encoded into $C_1$ code where for the chosen gate library, each non-transversal gate on $C_1$ must have the form of $U=C^{k}Z(\theta)$. Recall that Fig. \ref{fig:CkZ} shows the circuit for applying the non-transversal implementation of such a $U$ on $C_1$. In the nonuniform code concatenation approach \cite{nikahd2016non16}, only one quantum code, namely $C_2$, is used to encode the active qubits of $C_1$ in the second level of concatenation, where all of the gates in this circuit can be applied transversally on $C_2$. However, in this section we extend the nonuniformity of the code by allowing to use two codes $C_2$ and $C_3$ for encoding the active qubits. In other words, $q_t$ is encoded into $C_3$ whereas $C_2$ is used to encode the remaining active qubits. The non-active qubits can be left unencoded or encoded using either $C_1$ or $C_2$. As in the previous section, we refer to these cases as Case I, Case II and Case III, respectively. $C^{k}Z(\theta)$ must be transversal on $C_3$ and as stated in Section \ref{sec:proposed1}, it is only needed for $C_2$ to be a CSS code.

Now, a main challenge remains: How a CNOT gate can be applied fault-tolerantly between the codewords of different quantum codes, i.e., $C_2$ and $C_3$? Theorem \ref{theorem:PFT_CX} \cite{yoder2016universal21} answers this question.

\begin{theorem}
	For any two non-degenerate stabilizer codes, each of which either (a) possesses a complete set of fault-tolerant logical local Cliffords or (b) is CSS with a logical local Clifford gate, there exists a logical CNOT gate between them in both directions.
	\label{theorem:PFT_CX}
\end{theorem}

For two codes $C_2=[[n_2, 1, d_2]]$ and $C_3=[[n_3, 1, d_3]]$ which satisfy Theorem \ref{theorem:PFT_CX}, a logical CNOT gate can be applied as follows: Let $SX_{2}(SZ_{2})$ and $SX_{3}(SZ_{3})$ be the supports of the logical operator $X(Z)$ for $C_2$ and $C_3$, respectively, on which the corresponding logical operator only has $X(Z)$ and I. Connecting physical CNOT gates from $SZ_{2}$ to $SX_{3}$ in a round-robin fashion (considering all combinations) applies a logical CNOT gate between $C_2$ (as control) and $C_3$ (as target). This implementation is non-transversal. To make it fault-tolerant, the PFT method can be used. Based on this method, the circuit is broken into some $(min(d_2,d_3)-1)$-transversal pieces and the intermediate error corrections are inserted between them to correct errors, before they propagate to a set of non-correctable errors \cite{yoder2016universal21}. 

\begin{figure}
	\centering
	\includegraphics[trim=0in 0in 0in 0in,clip,width=0.9\columnwidth, height=0.38\textheight]{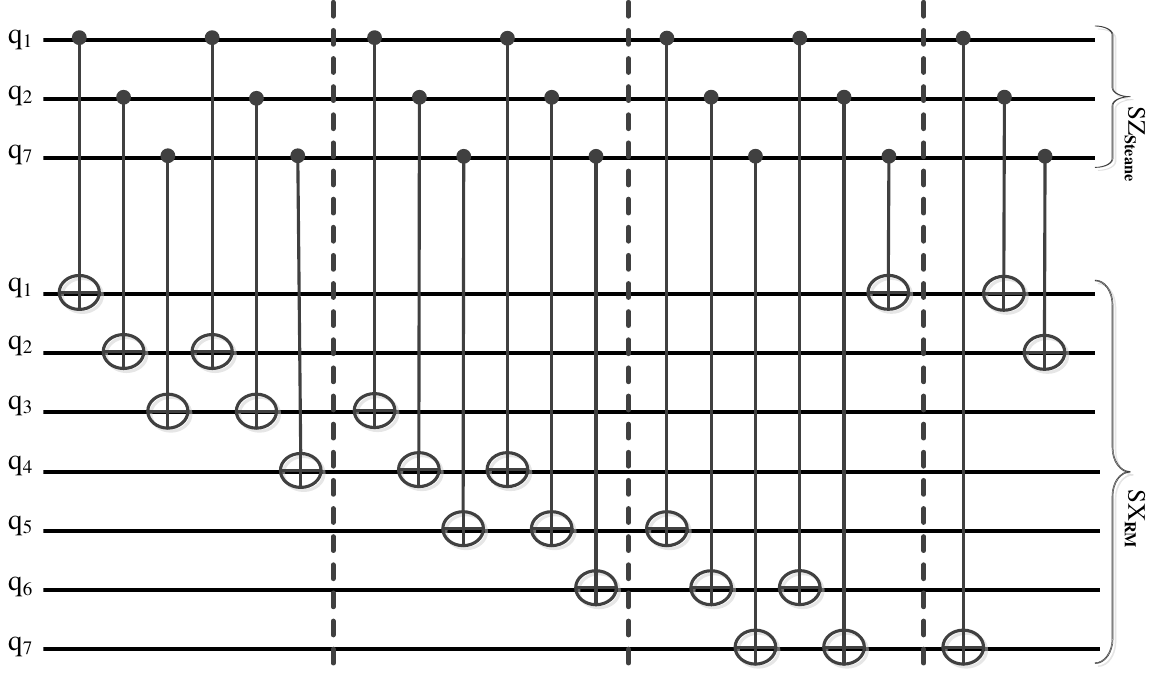}
	\caption{Round-robin implementation of a logical CNOT gate from a Steane code block to an RM code block which is pieceably fault-tolerant with four pieces.}
	\label{fig:CX_7to15_RR}
\end{figure}

\begin{figure}
	\centering
	\includegraphics[trim=0in 0in 0in 0in,clip,width=0.8\columnwidth, height=0.6\textheight]{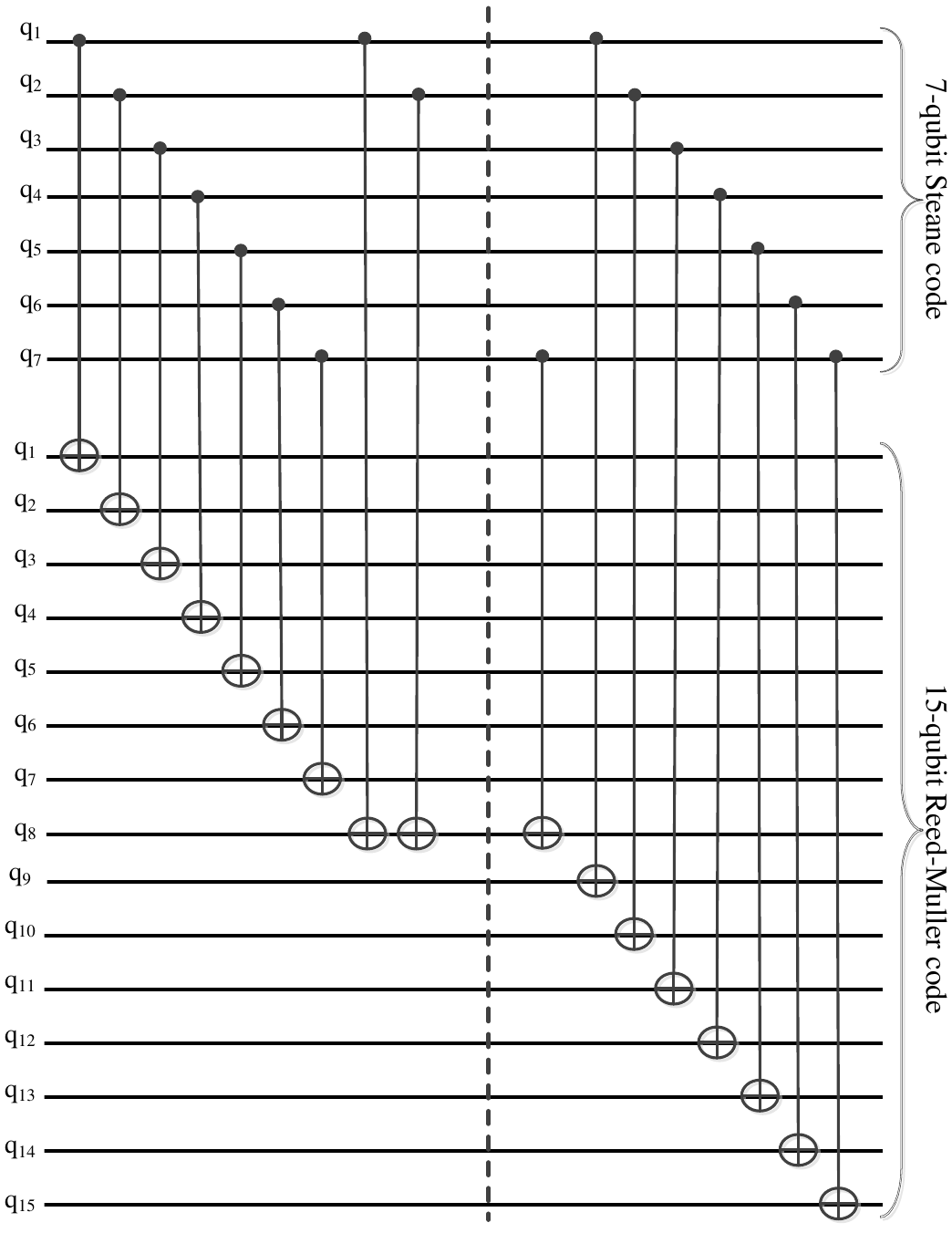}
	\caption{The proposed implementation of a logical CNOT gate from a Steane code block to an RM code block which is pieceably fault-tolerant with two pieces.}
	\label{fig:CX_7to15}
\end{figure}

As an example, Fig. \ref{fig:CX_7to15_RR} shows a round-robin circuit which applies a logical CNOT gate from Steane to 15-qubit Reed-Muller code. This circuit consists of 21 CNOTs in four pieces and thus, three intermediate error corrections and a final error correction are required. However, we found a more efficient circuit consisting of only 17 CNOTs with two pieces \footnote{The correctness of this circuit is verified using Quipu simulator \cite{garcia2014simulation} for all basis states.}. Fig. \ref{fig:CX_7to15} shows this circuit. 

\begin{figure}
	\centering
	\includegraphics[trim=0in 0in 0in 0in,clip,width=0.7\columnwidth, height=0.25\textheight]{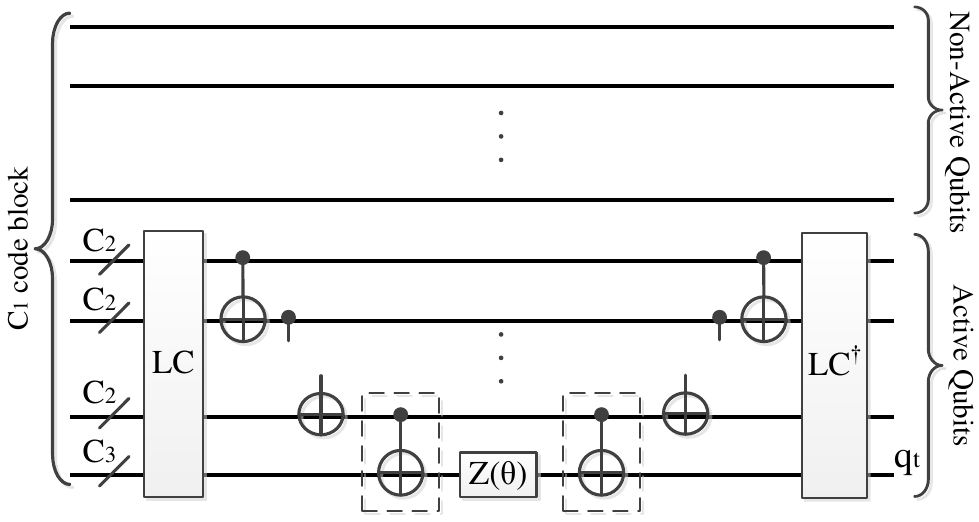}
	\caption{Fault-tolerant implementation of a logical $Z(\theta)$ gate based on the ENUCC approach.}
	\label{fig:ENUCC}
\end{figure}

A general schematic of the proposed approach for applying a $Z(\theta)$ gate is shown in Fig. \ref{fig:ENUCC}. The CNOT gates surrounded by dashed boxes should be applied based on the PFT method as they are applied on codewords of different codes, i.e., $C_2$ and $C_3$.

So far, fault-tolerant implementation of non-transversal gates on $C_1$ has been described. Note that these gates are not fully transversal and a single physical error on one of the active qubits can spread to another one. However, as all of the gates in the second level of concatenation are implemented fault-tolerantly, this error propagates to at most one physical qubit in each code block which can be corrected using error correction procedure in the second level of coding hierarchy. The ways in which the other gates are applied in ENUCC approach are similar to the HCC method as described in the previous section. 

In the next section we will elaborate more on the ENUCC method through some examples using the 5-qubit, Steane and 15-qubit Reed-Muller code ($RM$) codes. 

\subsection{ENUCC-based code examples using the Steane code as $C_1$} \label{subsec:steane_ENUCC}

In this section, the Steane code is considered as $C_1$ with \{$H$, $S=C^{0}Z(\frac{\pi}{2})$, $CZ=C^{1}Z(\pi)$, $T=C^{0}Z(\frac{\pi}{4})$\} as the universal gate set. As mentioned before, the Steane code is a CSS code with only non-transversal implementation for $T$ but $T$ is transversal on $RM$. Therefore, Steane and $RM$ can be selected as $C_2$ and $C_3$, respectively. This choice means that, $q_7$ ($q_t$) is encoded into $RM$, the remaining active qubits ($q_1$ and $q_2$) are encoded using Steane and the non-active qubits ($q_3$ to $q_6$) can be encoded using Steane or left unencoded. Based on whether the non-active qubits are left unencoded or encoded using Steane, a 33- or 57-qubit code is produced, respectively. 

\begin{figure}
	\centering
	\includegraphics[trim=0in 0in 0in 0in,clip,width=0.6\columnwidth]{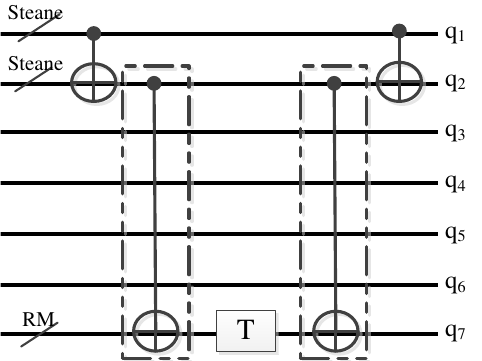}
	\caption{Fault-tolerant implementation of $T$ based on the ENUCC approach for the 33-qubit code.}
	\label{fig:steaneT_ENUCC}
\end{figure}

$S$ and CNOT gates are transversal on both Steane and $RM$ and are thus, fault-tolerant for the proposed concatenated codes. The logical $T$ gate can be applied fault-tolerantly as shown in Fig. \ref{fig:steaneT_ENUCC}, where the circuit depicted in Fig. \ref{fig:CX_7to15} is used to apply the CNOT gates surrounded by the dashed boxes. While $H$ is transversal on Steane, it is not transversal on $RM$. However, even though a single physical error on an $RM$ code block causes a logical error on that code block, this error can be corrected using error correction on $C_1$.

\subsection{ENUCC-based code examples using the 5-qubit code as $C_1$} \label{subsec:5qubitCode_ENUCC}

In this section, like Section \ref{subsec:5qubitCode_HCC}, the set M=\{$T=C^{0}Z(\frac{\pi}{4})$, $S=C^{0}Z(\frac{\pi}{2})$, $CZ=C^{1}Z(\pi)$\} along with $K$ are considered as the universal gate set. In the proposed codes in this section, a logical qubit is encoded into the 5-qubit code in the first level of concatenation. While the gates of $M$ are not transversal on the 5-qubit code, they belong to the class of $C^{k}Z(\theta)$ gates and are transversal on $RM$. Therefore, Steane and $RM$ can be selected as $C_2$ and $C_3$, respectively. Based on this code selection, $q_3$ ($q_t$) is encoded using $RM$ and $q_1$ and $q_5$ are encoded into the Steane code. The non-active qubits, $q_2$ and $q_4$, can be left unencoded or encoded using the 5-qubit code or Steane leading to a 31-, 39- or 43-qubit code, respectively. 

\begin{figure}
	\centering
	\includegraphics[trim=0in 0in 0in 0in,clip,width=0.6\columnwidth]{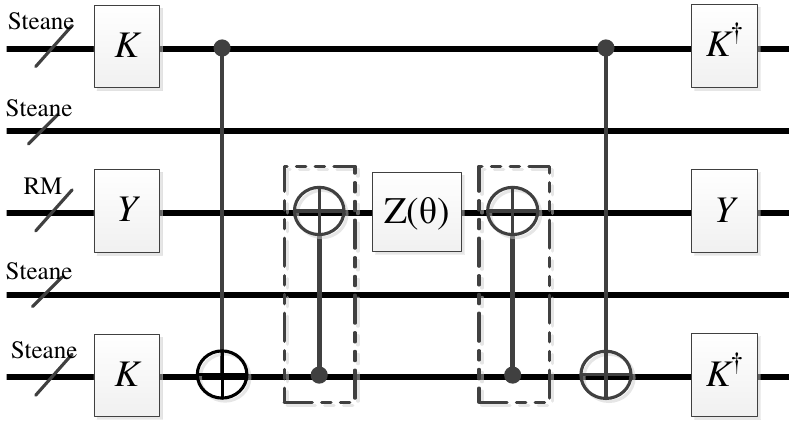}
	\caption{Fault-tolerant implementation of $Z(\theta)$ based on the ENUCC approach for the 43-qubit code.}
	\label{fig:5_qubitT_ENUCC}
\end{figure}

The gates of $M$ can be applied fault-tolerantly, as shown in Fig. \ref{fig:5_qubitT_ENUCC}, where the CNOT gates surrounded by the dashed boxes are applied using the circuit depicted in Fig. \ref{fig:CX_7to15}. The $K$ gate is non-transversal on $RM$. However, an erroneous $RM$ code block during application of this gate can be corrected using error correction on the 5-qubit code.

\section{Code analysis} \label{sec:codeAnalysis}
Straight concatenation of the two codes [[$n_1$,1,$d_1$]] and [[$n_2$,1,$d_2$]] leads to a code [[$n_1$$n_2$,1,$d_1d_2$]] \cite{gottesman1997stabilizer22}. However, the obtained distance ($d_1d_2$) may be reduced because of error propagation during application of a non-transversal gate. We refer to $d_1d_2$ as the overall distance of the concatenated code and use \emph{effective distance} to refer to the reduced distance. In this section, we analyze the proposed codes in terms of the overall and effective distance as well as their resource overhead. 

For the concatenated codes with fully encoded qubits in both levels of concatenation (Cases II and III), e.g., the 31-, 35- and 49-qubit HCC-based codes and 39-, 43- and 57-qubit ENUCC-based codes, the overall distance is 9. On the other hand, for the codes with unencoded non-active qubits (Case I), i.e., the 23- and 25-qubit HCC-based codes and 31- and 33-qubit ENUCC-based codes, the overall distance is 5 as deduced in \cite{chamberland2016architectural23} for the 49-qubit nonuniform concatenated code \cite{nikahd2016non16}.

The effective distance of the proposed codes varies for different gates. Table \ref{tab:comparison} compares the effective distance of the proposed codes with their corresponding uniform (UCC) \cite{jochym2014using15} and nonuniform (NUCC) \cite{nikahd2016non16} concatenated codes for the gates of their selected universal gate sets.

\begin{table}[h]
	\begin{center}
		\begin{minipage}{\textwidth}
			\caption{Comparison of the codes based on various code concatenation approaches i.e., UCC, NUCC, HCC and ENUCC, in terms of the number of qubits, effective distance for different gates and overall distance.}\label{tab:comparison}
			\begin{tabular*}{\textwidth}{@{\extracolsep{\fill}}lccccccccc@{\extracolsep{\fill}}}
				\toprule%
				\multirow{2}{*}{Method} & \multirow{2}{*}{$C_1$} & \multirow{2}{*}{Case} & \multirow{2}{*}{\#qubits} & \multicolumn{5}{@{}@{}c@{}@{}}{Effective Dist.} &  \multirow{2}{*}{Overall Dist.} \\\cmidrule{5-9} 
				& & & & H & K & T & S & CZ & \\
				\midrule
				\multirow{2}{*}{UCC} & Steane & - & 105 & 3 & 3 & 3 & 9 & 9 & 9\\\cmidrule{2-10}
				& 5-qubit & - & 75 & - & 3 & 3 & 3 & 3 & 9\\ 
				\midrule
				\multirow{4}{*}{NUCC} & \multirow{2}{*}{Steane} & I & 49 & 3 & 3 & 3 & 5 & 5 & 5\\
				& & II $\equiv$ III & 73 & 3 & 3 & 3 & 9 & 9 & 9\\\cmidrule{2-10}
				& \multirow{2}{3.2em}{5-qubit} & I & 47 & - & 3 & 3 & 3 & 3 & 5\\
				& & II & 55 & - & 3 & 3 & 3 & 3 & 9\\
				\midrule				
				\multirow{5}{*}{HCC} & \multirow{2}{*}{Steane} & I & 25 & 5 & 5 & 3 & 5 & 5 & 5\\
				& & II $\equiv$ III & 49 & 9 & 9 & 3 & 9 & 9 & 9\\\cmidrule{2-10}
				& \multirow{3}{*}{5-qubit} & I & 23 & - & 5 & 3 & 3 & 3 & 5\\
				& & II & 31 & - & 9 & 3 & 3 & 3 & 9\\
				& & III & 35 & 9 & 9 & 3 & 9 & 3 & 9\\
				\midrule				
				\multirow{5}{*}{ENUCC} & \multirow{2}{*}{Steane} & I & 33 & 3 & 3 & 3 & 5 & 5 & 5\\
				& & II $\equiv$ III & 57 & 7 & 7 & 3 & 9 & 9 & 9\\\cmidrule{2-10}
				& \multirow{3}{3.2em}{5-qubit} & I & 31 & - & 3 & 3 & 3 & 3 & 5\\
				& & II & 39 & - & 7 & 3 & 3 & 3 & 9\\
				& & III & 43 & 7 & 7 & 3 & 7 & 3 & 9\\
				\botrule
			\end{tabular*}
		\end{minipage}
	\end{center}
\end{table}

For the shared transversal gates in both levels of concatenation, no error is propagated in the code blocks and thus, for these gates, the effective distance of a code will be equal to its overall distance. The gates with the effective distance of 5 and 9 in Table \ref{tab:comparison} are examples of such gates. 

The $H$ and $K$ gates are non-transversal on $RM$. Consequently, for these gates, the effective distance of the uniform and nonuniform codes, which have three or more code blocks of $RM$ in their second level of concatenation, is reduced from 9 or 5 to 3. For the ENUCC-based codes, which have only one code block of $RM$, the effective distance is reduced from 5 to 3 or from 9 to 7. For example, for the 57-qubit code, the worst case is when two errors on a Steane code block and one error on the $RM$ code block occur that none of them can be corrected using error correction in the second level. In this case, the code block that is complementary to these two errors is identified when the $C_1$ stabilizers are measured. However, since no error had been identified on that code block in the second level, the complementary errors can be clearly detected. Although adding one additional physical error on another Steane code block of the second level makes it ambiguous to correctly identify the erroneous qubits. 

\begin{figure}
	\centering
	\includegraphics[trim=0in 0in 0in 0in,clip,width=0.42\columnwidth, height=0.22\textheight]{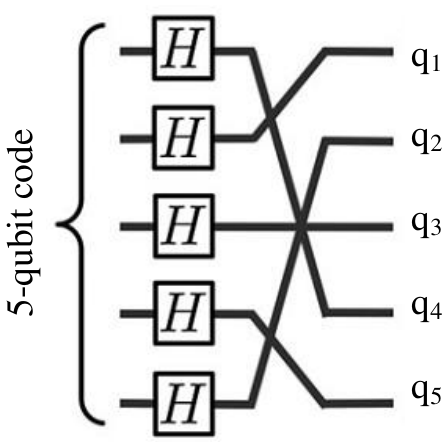}
	\caption{Transversal implementation of $H$ on the 5-qubit code by permuting qubits. Note that $q_3$ is not permuted.}
	\label{fig:5_qubitH_permute}
\end{figure}

Note that the $H$ gate is applicable for the 35- and 43-qubit codes with the effective distance of 9 and 7, respectively. This is because $H$ is transversal on the 5-qubit code by permuting qubits as shown in Fig. \ref{fig:5_qubitH_permute} \cite{yoder2016universal21}. This permutation is applicable for these two codes, as the permuted qubits during application of $H$, i.e., $q_1$, $q_2$, $q_4$ and $q_5$, are encoded blocks of the same code, i.e., Steane code. However, applying $H$ with permutation for the other proposed codes based on the 5-qubit code destroys the code structure as it permutes the code blocks that are encoded using different codes in the second level of concatenation. Generally, a transversal gate with permutation on $C_1$ is not applicable to nonuniform concatenated codes \cite{nikahd2016non16} unless it permutes only the encoded blocks of the same code. 

For the non-transversal gates on $C_1$, all of the code examples have the effective distances of 3. In this case, a single physical error on a qubit of $C_1$ propagates to at most one physical error on each active qubit and these errors can be corrected using error correction procedure on $C_2$ code blocks. \{$S$, $CZ$, $T$\} and \{$T$\} are such gates for the proposed codes based on the 5-qubit and Steane codes, respectively. However, for the $S$ gate, the effective distances of the 35- and 43-qubit codes are 9 and 7, respectively. This is because $S$ can also be applied as $KH$. It should be noted that $CCZ$ which is equal to $C^{2}Z(\pi)$ can be applied fault-tolerantly for the proposed codes as its implementation on the Steane and 5-qubit codes has the same structure as $T$ and it is also transversal on $RM$.

The proposed code examples outperform the UCC- and NUCC-based codes proposed in \cite{jochym2014using15} and \cite{nikahd2016non16} in terms of both number of qubits and effective distance. For example, the counterparts of the 105-qubit UCC-based code based on the NUCC, ENUCC and HCC approaches are the 73-, 57- and 49-qubit codes, respectively. 
All of these codes have the same overall distance and also, the same effective distances for $S$, $CZ$ and $T$ gates. For the $H$ and $K$ gates, the effective distance of 105- and 73-qubit codes is 3, while the ENUCC and HCC approaches increase this distance to 7 and 9, respectively. This result becomes more valuable by the fact that the threshold of the 49- and 105-qubit concatenated codes are limited by the application of logical $H$ gate \cite{chamberland2016architectural23,chamberland2016thresholds22}.

This raise in effective distance is achieved at the cost of more resource overhead caused by applying $T$ gate. This resource overhead is incured by MSD, code switching or PFT. However, for all operations other than $T$, the 49- and 57-qubit codes have less resource overhead. This is because these codes are smaller than the 73- and 105-qubit codes and also have fewer $RM$ code blocks.


The 49-qubit HCC-based code is smaller than the 57-qubit ENUCC-based code with no $RM$ code blocks in its structure and thus, has less resource overhead for applying $H$, $K$, $S$, $CZ$ and error correction procedure. The only chance of the 57-qubit code to surpass the 49-qubit code is the resource overhead of $T$. However, our analysis shows that the 57-qubit code falls behind the 49-qubit code in terms of resource overhead with applying this gate, as shown in Table \ref{tab:comparison2}. Therefore, the 57-qubit code has no advantage over the 49-qubit code. Indeed, a pieceable CNOT gate between Steane and $RM$ codes is more resource intensive than an MSD-based $T$ gate application on Steane. Nevertheless, the ENUCC method may surpass the HCC appraoch for other code combinations.

For comparison with MSD, the two-level concatenated Steane code (equivalent to the 49-qubit HCC-based code) is used where the MSD approach is utilized to apply $T$ on this code. Table \ref{tab:comparison2} shows that the HCC approach significantly reduces the resource overhead and delay for applying $T$ gate in comparison with MSD. This improvement is achieved at the cost of reducing the effective distance of the code from 9 to 3, for applying this gate. 

The circuits used for implementing each operation and their resource overheads' formulations are given in Appendix \ref{sec:appendix}. The parameters used for delay calculations are $T_{1-qubit}=10 \mu s$, $T_{2-qubit}=100 \mu s$, $T_{M}=100 \mu s$.

\begin{table}[h]
	\begin{center}
		\begin{minipage}{\textwidth}
			\caption{Resource overhead and delay comparison of $T$ gate for the 49-qubit MSD- and HCC-based codes based on two-level concatenated Steane and the 57-qubit ENUCC codes.}\label{tab:comparison2}
			\begin{tabular*}{\textwidth}{@{\extracolsep{\fill}}lcccccc@{\extracolsep{\fill}}}
				\toprule%
				\multirow{2}{*}{Approach} & \multicolumn{3}{@{}c@{}}{Resource Overhead} & \multirow{2}{*}{Delay($\mu s$)} \\\cmidrule{2-4} 
				 & 1-qubit & 2-qubit & M &  \\
				\midrule
				MSD & 361     & 631     & 70  & 5820\\
				ENUCC & 243     & 482     & 228 & 2620\\
				HCC & 53      & 106     & 14  & 1830\\ 
				ENUCC/MSD  & 0.67 & 0.76 & 3.25 & 0.58\\ 	
				HCC/MSD  & 0.15 & 0.17 & 0.20 & 0.32\\
				\botrule
			\end{tabular*}
		\end{minipage}
	\end{center}
\end{table}

It is worth noting that the our 25-qubit HCC-based code is explored in \cite{lin2020concatenated} in details\footnote{The preprint version of our article has been available in arXiv sine 2017 \cite{nikahd2017low}.}, where a pieceable-based code switching approach is used for fault-tolerant application of $T$ gate. They found the pseudothreshold of the 25-qubit code to be $3.93\times10^{-4}$.

\section{\label{sec:conclusion}Conclusion}
In this study, we proposed two low-overhead approaches for universal FTQC, namely HCC and ENUCC. The HCC method combines code concatenation with code switching, MSD or PFT schemes. ENUCC uses two codes, $C_2$ and $C_3$, for encoding active qubits by allowing to apply fault-tolerant CNOT gates between codewords of different codes, i.e., $C_2$ and $C_3$. 

The proposed approaches were described based on the 5-qubit and Steane codes in two levels of concatenation as examples which lead to the 25-, 49-, 23-, 31- and 35-qubit HCC-based and 33-, 57-, 31-, 39- and 43-qubit ENUCC-based codes. In comparison to UCC and NUCC approaches, the proposed codes are smaller and also increase the effective distance and reduce the resource overhead for all of the logical operations, except for $T$. These improvements are achieved at the cost of more resource overhead for applying this gate. Furthermore, these approaches significantly reduce the resource overhead in comparison with MSD at the cost of reducing the effective distance of the concatenated code for implementing non-transversal gates.

For the proposed code examples, the ENUCC-based codes have no advantages over the HCC-based codes. However, it may outperform HCC for other code combinations. 

\bmhead{Acknowledgments}
The authors acknowledge the financial support by the INSF.

\section*{Declarations}

\subsection*{Funding}
This study was supported by a grant from Iran National Science Foundation.

\subsection*{Conflict of interest/Competing interests}
The authors have no conflicts of interest to declare.

\subsection*{Ethics approval}
Not applicable.

\subsection*{Consent to participate}
Not applicable.

\subsection*{Consent for publication}
Not applicable.

\subsection*{Availability of data and materials}
No dataset is used in this article. The article contains almost all data and materials needed to replicate the results (in the main text or the appendix).

\subsection*{Code availability}
Not applicable.

\begin{appendices}
	
	\section{\label{sec:appendix}Circuits used for resource comparison}
	In this appendix the circuits used for resource comparison in Section \ref{sec:codeAnalysis} are depicted and formulated with the following assumptions and notations.
	\begin{itemize}
		\item We assume that any postselections always succeed. 
		\item Pauli corrections are not considered as they can be handeled using Pauli frame without explicit application. However, it is assumed that the non-Pauli corrections always need to be applied.
		\item $n\times U$ represents $n$ times executions of $U$ operation but in parallel. 
		\item $U_1 + U_2$ stands for serial application of $U_1$ and $U_2$.
		\item Parallel application of two operations $U_1$ and $U_2$ is shown by $U_1 \vert \vert U_2$.
		\item $Zero_n$ represents preparation of the logical $ \vert 0 \rangle $ state for an \textit{n}-qubit code. Preparing a single physical qubit into the $ \vert 0 \rangle$ state is shown by $Zero_1$. 
		\item $Plus_n$ represents preparation of the $ \vert + \rangle$ state for an \textit{n}-qubit code.
		\item $QEC_n$ stands for error correction procedure for an \textit{n}-qubit code.
		\item $M_x(M_z)$ stands for measuring a qubit in $x(z)$ basis.
	\end{itemize}
	
	\subsection{T gate implementation on the 49-qubit HCC-based code}
	
	Fig. \ref{fig:steaneT_HCC} shows the circuit for applying $T$ on the 49-qubit HCC-based code by using the MSD method in the second level of concatenation. The resource overhead of this gate can be formulated as:
	\begin{equation}
		\begin{split}
			T: 7\times CNOT+7\times CNOT+ T_{MSD7} \\+ 7\times CNOT+7\times CNOT.
		\end{split}
	\end{equation}
	
	\subsection{MSD-based implementation of T gate on Steane code}
\begin{figure}
		\centering
		\includegraphics[trim=0in 0in 0in 0in,clip,width=0.4\columnwidth, height=0.1\textheight]{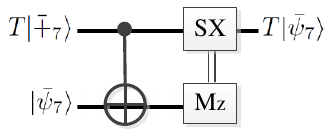}
		\caption{The $T$ gate implementation on a Steane code block using the magic state $T \vert \bar{+}_7 \rangle$.}
		\label{fig:TMSD}
	\end{figure}

	The MSD-based $T$ gate can be applied on a Steane code block as shown in Fig. \ref{fig:TMSD} using the magic state $T \vert \bar{+}_7\rangle$. Therefore, the resource overhead of this gate can be formulated as: 
	\begin{equation}
		\begin{split}
			T_{MSD7}: T \vert \bar{+}_7 \rangle+7\times CNOT+7\times M_z+7\times S^\dag.
		\end{split}
	\end{equation}
	
\begin{figure}
		\centering
		\includegraphics[trim=0in 0in 0in 0in,clip,width=0.6\columnwidth, height=0.18\textheight]{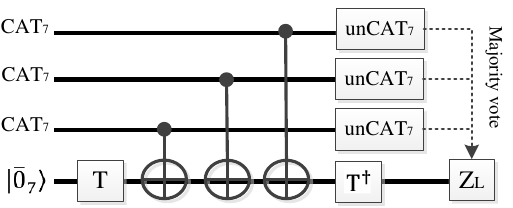}
		\caption{Fault tolerant circuit to prepare $T \vert \bar{+}_7 \rangle$ magic state.}
		\label{fig:TPLUS}
	\end{figure}

	The magic state $T \vert \bar{+}_7\rangle$ can be prepared using the circuit shown in Fig. \ref{fig:TPLUS}, without the need for distillation \cite{ahsan2015architecture7}\cite{nielsen2010quantum3}. The resource overhead of this operation may be calculated using the following relation:
	\begin{equation}
		\begin{split}
			&T \vert \bar{+}_7\rangle: (3\times CAT_7 \vert \vert (Zero_7+ 7\times T))\\&+7\times CNOT+unCAT_7 \vert \vert (7\times CNOT+ \\&unCAT_7 \vert \vert (7\times CNOT+(unCAT_7 \vert \vert 7\times T))),
		\end{split}
	\end{equation}
	
	where $CAT_7$ and $unCAT_7$ refer to the 7-qubit CAT state preparation and its decoding. For two-level concatenated Steane code, the MSD-based $T$ gate can be applied similarly, except that the 7-qubit logical blocks and gates should be replaced by the 49-qubit ones.
	
	\subsection{7-qubit CAT state preparation and its decoding}
\begin{figure}
		\centering
		\includegraphics[trim=0in 0in 0in 0in,clip,width=0.7\columnwidth, height=0.22\textheight]{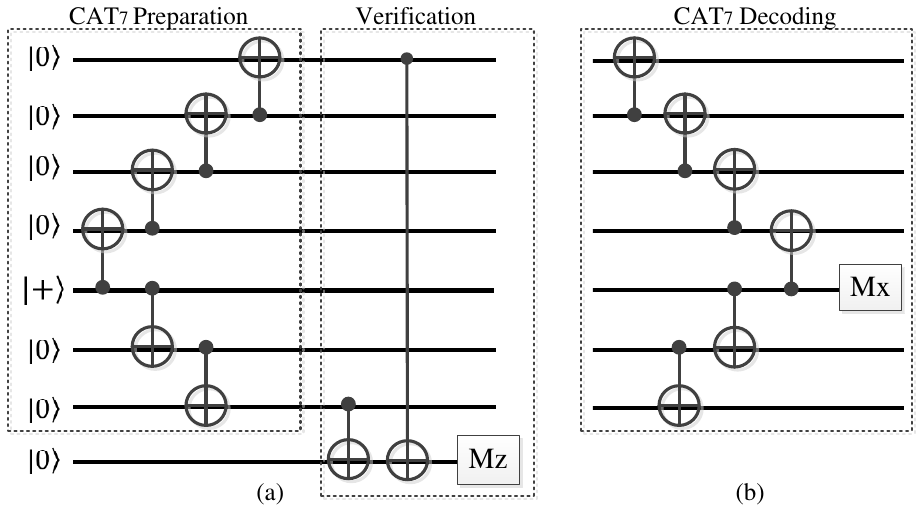}
		\caption{a) $CAT_7$ state preparation and b) $CAT_7$ state decoding.}
		\label{fig:cat7}
	\end{figure}

	Fig. \ref{fig:cat7} shows the circuits for preparing the 7-qubit CAT state ($CAT_7$) and its decoding ($unCAT_7$). The resource overhead of these implementations can obtained by: 
	\begin{equation}
		\begin{split}
			&CAT_7: (7\times Zero_1 \vert \vert Plus_1)+CNOT+2\times CNOT\\
			&+2\times CNOT+2\times CNOT+CNOT+M_z,
		\end{split}
	\end{equation}
	
	\begin{equation}
		\begin{split}
			&unCAT_7: CNOT+2\times CNOT+2\times CNOT+\\&CNOT+M_z.
		\end{split}
	\end{equation}
	
	\subsection{T gate implementation on the 57-qubit ENUCC-based code}
	Fig. \ref{fig:steaneT_ENUCC} shows the circuit for applying $T$ on the 33-qubit ENUCC-based code by using the PFT method in the second level of coding hierarchy. The resource overhead of this gate for both the 33- and 57-qubit ENUCC-based codes are the same and can be found as:
	\begin{equation}
		\begin{split}
			T: 7\times CNOT+PCNOT_{7-15}+15\times T\\+PCNOT_{7-15}+7\times CNOT.
		\end{split}
	\end{equation}
	
	\subsection{Pieceable CNOT from Steane to RM}
	Fig. \ref{fig:CX_7to15} shows a pieceable implementation of CNOT from Steane to $RM$. The resource overhead of this implementation can be formulated as: 
	\begin{equation}
		\begin{split}
			&PCNOT_{7-15}: 7\times CNOT+CNOT+CNOT\\
			&+(QEC_7^z \vert \vert QEC_{15}^z)+CNOT+7\times CNOT\\
			&+(QEC_7 \vert \vert QEC_{15}).
		\end{split}
	\end{equation}
	
	\subsection{Quantum error correction on Steane code}
\begin{figure}
		\centering
		\includegraphics[trim=0in 0in 0in 0in,clip,width=0.7\columnwidth, height=0.16\textheight]{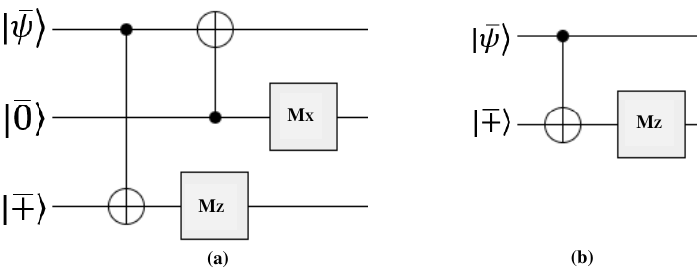}
		\caption{a) Quantum error correction based on the Steane's method \cite{steane1997active}, b) The measurement of only the Z stabilizers of a CSS code.}
		\label{fig:QEC}
	\end{figure}

	Quantum error correction based on the Steane's syndrome extraction method \cite{steane1997active} is shown in Fig. \ref{fig:QEC}. For the 7-qubit Steane code, this operation is formulated as: 
	\begin{equation}
		\begin{split}
			QEC_7: (Zero_7 \vert \vert Plus_7)+7\times CNOT\\+(7\times CNOT \vert \vert 7\times M_z)+7\times M_x.
		\end{split}
	\end{equation}
	
	\subsection{Logical $\vert \overline{0}\rangle$ and $\vert \overline{+}\rangle$ preparation for the Steane code}
	The circuit suggested in \cite{goto2016minimizing} is used for logical $\vert \overline{0}_7\rangle$ and $\vert \overline{+}\rangle$ preparation and can be computed using the following relations:
	\begin{equation}
		\begin{split}
			Zero_7: (5\times Zero_1 \vert \vert 3\times Plus_1) +3\times CNOT+3\times \\
			CNOT+3\times CNOT+CNOT+CNOT+M_z,
		\end{split}
	\end{equation}
	\begin{equation}
		\begin{split}
			Plus_7:(3\times Zero_1 \vert \vert 5\times Plus_1) +3\times CNOT+3\times \\
			CNOT+3\times CNOT+CNOT+CNOT+M_x.
		\end{split}
	\end{equation}
	
	
	\begin{figure}
		\centering
		\includegraphics[trim=0in 0in 0in 0in,clip,width=0.6\columnwidth, height=0.13\textheight]{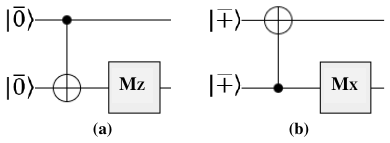}
		\caption{Fault-tolerant preparation of an encoded state a) $\vert \overline{0}\rangle$ and b) $\vert \overline{+}\rangle$ for a CSS code.}
		\label{fig:prepFT}
	\end{figure}
	
	\subsection{Quantum error correction on $RM$ code}
	The resource overhead of error correction on an $RM$ code block is obtained by:
	\begin{equation}
		\begin{split}
			QEC_{15}: (Zero_{15} \vert \vert  Plus_{15})+15\times CNOT\\
			+(15\times CNOT \vert \vert 15\times M_z)+15\times M_x,
		\end{split}
	\end{equation}
	\begin{equation}
		\begin{split}
			QEC_{15}^z: Plus_{15}+15\times CNOT+15\times M_z.
		\end{split}
	\end{equation}
	
	\subsection{Logical $\vert \overline{0}\rangle$ and $\vert \overline{+}\rangle$ preparation for the $RM$ code}
	
	A fault-tolerant (FT) $\vert \overline{0}_{15}\rangle$/$\vert \overline{+}_{15}\rangle$ can be prepared by using two non-FT $\vert \overline{0}_{15}\rangle$/$\vert \overline{+}_{15}\rangle$ states as shown in Fig. \ref{fig:prepFT}. The circuit for non-FT preparation of $\vert \overline{0}_{15}\rangle$/$\vert \overline{+}_{15}\rangle$, denoted by $NFTZero_{15}/NFTPlus_{15}$, is taken from \cite{chamberland2016architectural23}.
	
	\begin{equation}
		Zero_{15}: 2\times NFTZero_{15} +15\times CNOT+15 \times M_z,
	\end{equation}
	\begin{equation}
		Plus_{15}: 2\times NFTPlus_{15} +15\times CNOT+15 \times M_x,
	\end{equation}
	\begin{equation}
		\begin{split}
			NFTZero_{15}: (11\times Zero_1\vert \vert 4\times Plus_1)+4\times \\
			CNOT+4\times CNOT+4\times CNOT+4\times \\
			CNOT+3\times CNOT+3\times CNOT,
		\end{split}
	\end{equation}
	\begin{equation}
		\begin{aligned}
			NFTPlus_{15}: (10\times Zero_1 \vert \vert  5\times Plus_1)+5\times \\
			CNOT+5\times CNOT+5\times CNOT+5\times \\
			CNOT+3\times CNOT+2\times CNOT.
		\end{aligned}
	\end{equation}

\end{appendices}

\bibliography{HENUCCIJTP}

\providecommand{\noopsort}[1]{}\providecommand{\singleletter}[1]{#1}%
\begin{thebibliography}{10}
\providecommand{\url}[1]{{#1}}
\providecommand{\urlprefix}{URL }
\providecommand{\doi}[1]{\url{https://doi.org/#1}}
\bibcommenthead

\bibitem{ahsan2015architecture7}
M.~Ahsan, Architecture framework for trapped-ion quantum computer based on
  performance simulation tool.
\newblock Ph.D. thesis, Duke University (2015)

\bibitem{shor1994algorithms1}
P.W. Shor, in \emph{Foundations of Computer Science, 1994 Proceedings., 35th
  Annual Symposium on} (IEEE, 1994), pp. 124--134

\bibitem{zalka1998efficient2}
C.~Zalka, Efficient simulation of quantum systems by quantum computers.
\newblock Fortschritte der Physik \textbf{46}(6-8), 877--879 (1998)

\bibitem{nielsen2010quantum3}
M.A. Nielsen, I.L. Chuang, \emph{Quantum computation and quantum information}
  (Cambridge university press, 2010)

\bibitem{unruh1995maintaining4}
W.G. Unruh, Maintaining coherence in quantum computers.
\newblock Phys.\ Rev. A \textbf{51}(2), 992 (1995)

\bibitem{mazzola2010sudden5}
L.~Mazzola, J.~Piilo, S.~Maniscalco, Sudden transition between classical and
  quantum decoherence.
\newblock Physical review letters \textbf{104}(20), 200,401 (2010)

\bibitem{metodi2006quantum6}
T.S. Metodi, F.T. Chong, Quantum computing for computer architects.
\newblock Synthesis Lectures in Computer Architecture \textbf{1}(1), 1--154
  (2006)

\bibitem{shor1996fault9}
P.W. Shor, in \emph{Foundations of Computer Science, 1996. Proceedings., 37th
  Annual Symposium on} (IEEE, 1996), pp. 56--65

\bibitem{anderson2014fault10}
J.T. Anderson, G.~Duclos-Cianci, D.~Poulin, Fault-tolerant conversion between
  the steane and reed-muller quantum codes.
\newblock Physical review letters \textbf{113}(8), 080,501 (2014)

\bibitem{eastin2009restrictions11}
B.~Eastin, E.~Knill, Restrictions on transversal encoded quantum gate sets.
\newblock Physical review letters \textbf{102}(11), 110,502 (2009)

\bibitem{bravyi2005universal12}
S.~Bravyi, A.~Kitaev, Universal quantum computation with ideal clifford gates
  and noisy ancillas.
\newblock Physical Review A \textbf{71}(2), 022,316 (2005)

\bibitem{paetznick2013universal17}
A.~Paetznick, B.W. Reichardt, Universal fault-tolerant quantum computation with
  only transversal gates and error correction.
\newblock Physical review letters \textbf{111}(9), 090,505 (2013)

\bibitem{bombin2015gauge18}
H.~Bomb{\'\i}n, Gauge color codes: optimal transversal gates and gauge fixing
  in topological stabilizer codes.
\newblock New Journal of Physics \textbf{17}(8), 083,002 (2015)

\bibitem{yoder2016universal21}
T.J. Yoder, R.~Takagi, I.L. Chuang, Universal fault-tolerant gates on
  concatenated stabilizer codes.
\newblock Physical Review X \textbf{6}(3), 031,039 (2016)

\bibitem{stephens2008asymmetric13}
A.M. Stephens, Z.W.E. Evans, S.J. Devitt, L.C.L. Hollenberg, Asymmetric quantum
  error correction via code conversion.
\newblock Physical Review A \textbf{77}(6), 062,335 (2008)

\bibitem{choi2015dual14}
B.S. Choi, Dual-code quantum computation model.
\newblock Quantum Information Processing \textbf{14}(8), 2775--2818 (2015)

\bibitem{jochym2014using15}
T.~Jochym-O’Connor, R.~Laflamme, Using concatenated quantum codes for
  universal fault-tolerant quantum gates.
\newblock Physical review letters \textbf{112}(1), 010,505 (2014)

\bibitem{nikahd2016non16}
E.~Nikahd, M.~Sedighi, M.~Saheb~Zamani, Nonuniform code concatenation for
  universal fault-tolerant quantum computing.
\newblock Phys. Rev. A \textbf{96}, 032,337 (2017).
\newblock \doi{10.1103/PhysRevA.96.032337}.
\newblock \urlprefix\url{https://link.aps.org/doi/10.1103/PhysRevA.96.032337}

\bibitem{fowler2012surface19}
A.G. Fowler, M.~Mariantoni, J.M. Martinis, A.N. Cleland, Surface codes: Towards
  practical large-scale quantum computation.
\newblock Physical Review A \textbf{86}(3), 032,324 (2012)

\bibitem{chamberland2019fault}
C.~Chamberland, A.W. Cross, Fault-tolerant magic state preparation with flag
  qubits.
\newblock Quantum \textbf{3}, 143 (2019)

\bibitem{chamberland2020very}
C.~Chamberland, K.~Noh, Very low overhead fault-tolerant magic state
  preparation using redundant ancilla encoding and flag qubits.
\newblock npj Quantum Information \textbf{6}(1), 1--12 (2020)

\bibitem{haah2018codes}
J.~Haah, M.B. Hastings, Codes and protocols for distilling $ t $, controlled-$
  s $, and toffoli gates.
\newblock Quantum \textbf{2}, 71 (2018)

\bibitem{hastings2018distillation}
M.B. Hastings, J.~Haah, Distillation with sublogarithmic overhead.
\newblock Physical review letters \textbf{120}(5), 050,504 (2018)

\bibitem{yoder2018practical}
T.J. Yoder, Practical fault-tolerant quantum computation.
\newblock Ph.D. thesis, Massachusetts Institute of Technology (2018)

\bibitem{oskin2002practical20}
M.~Oskin, F.T. Chong, I.L. Chuang, A practical architecture for reliable
  quantum computers.
\newblock Computer \textbf{35}(1), 79--87 (2002)

\bibitem{quan2018fault}
D.X. Quan, L.L. Zhu, C.X. Pei, B.C. Sanders, Fault-tolerant conversion between
  adjacent reed--muller quantum codes based on gauge fixing.
\newblock Journal of Physics A: Mathematical and Theoretical \textbf{51}(11),
  115,305 (2018)

\bibitem{beverland2021cost}
M.E. Beverland, A.~Kubica, K.M. Svore, Cost of universality: A comparative
  study of the overhead of state distillation and code switching with color
  codes.
\newblock PRX Quantum \textbf{2}(2), 020,341 (2021)

\bibitem{garcia2014simulation}
H.J. Garcia, I.L. Markov, Simulation of quantum circuits via stabilizer frames.
\newblock IEEE Transactions on Computers \textbf{64}(8), 2323--2336 (2014)

\bibitem{gottesman1997stabilizer22}
D.~Gottesman, Stabilizer codes and quantum error correction.
\newblock arXiv preprint quant-ph/9705052  (1997)

\bibitem{chamberland2016architectural23}
C.~Chamberland, T.~Jochym-O'Connor, R.~Laflamme, Overhead analysis of universal
  concatenated quantum codes.
\newblock Physical Review A \textbf{95}(2), 022,313 (2017)

\bibitem{chamberland2016thresholds22}
C.~Chamberland, T.~Jochym-O'Connor, R.~Laflamme, Thresholds for universal
  concatenated quantum codes.
\newblock Physical review letters \textbf{117}(1), 010,501 (2016)

\bibitem{lin2020concatenated}
C.~Lin, G.~Yang, Concatenated pieceable fault-tolerant scheme for universal
  quantum computation.
\newblock Physical Review A \textbf{102}(5), 052,415 (2020)

\bibitem{nikahd2017low}
E.~Nikahd, M.S. Zamani, M.~Sedighi, Low-overhead code concatenation approaches
  for universal quantum computation.
\newblock arXiv preprint arXiv:1707.00981  (2017)

\bibitem{steane1997active}
A.M. Steane, Active stabilization, quantum computation, and quantum state
  synthesis.
\newblock Physical Review Letters \textbf{78}(11), 2252 (1997)

\bibitem{goto2016minimizing}
H.~Goto, Minimizing resource overheads for fault-tolerant preparation of
  encoded states of the steane code.
\newblock Scientific reports \textbf{6}, 19,578 (2016)

\end{thebibliography}


\end{document}